# Quasi-1D graphene superlattices formed on high index surfaces

*Chenfang Lin[1†], Xiangqian Huang[1†], Fen Ke[1†], Chenhao Jin[1], Nai Tong[1], Xiuli Yin[1], Lin Gan[3], Xuefeng Guo[3], Ruguang Zhao[1], Weisheng Yang[1], Enge Wang[2,4] and Zonghai Hu[1,2]\**

[1]State Key Lab for Mesoscopic Physics, School of Physics, Peking University, Beijing 100871, China

[2]Collaborative Innovation Center for Quantum Matter, Beijing, China

[3]College of Chemistry and Molecular Engineering, Peking University, Beijing 100871, China

[4]International Center for Quantum Materials, School of Physics, Peking University, Beijing, China

We report preparation of large area quasi-1D monolayer graphene superlattices on a prototypical high index surface Cu(410)-O and characterization by Raman spectroscopy, Auger electron spectroscopy (AES), low energy electron diffraction (LEED), scanning tunneling microscopy (STM) and scanning tunneling spectroscopy (STS). The periodically stepped substrate gives a 1D modulation to graphene, forming a superlattice of the same super-periodicity. Consequently the moiré pattern is also quasi-1D, with a different periodicity. Scanning tunneling spectroscopy measurements revealed new Dirac points formed at the superlattice Brillouin zone boundary as predicted by theories.

[†] These authors contributed equally to this work.

\* zhhu@pku.edu.cn

# I. INTRODUCTION

Tuning graphene properties is one focus of current materials research. Among various proposed methods to tune graphene properties to meet certain needs, applying periodic potentials has many fascinating consequences such as deforming the Dirac cones, producing extra Dirac points, renormalizing the quasiparticle group velocity [1-10]. Obtaining various periodic patterns on graphene is thus a key issue. It remains challenging to make periodic patterns with a feature size of several nanometers through nano-fabrication. Alternatively, forming graphene superlattices on appropriate substrates is a natural route to realize periodic potentials. In fact, moiré patterns resulting from the lattice mismatch and/or rotation between graphene and the substrate are a type of commonly observed graphene superlattices. Up to now most work on graphene superlattices focuses on low index surfaces [4-8, 11-23] and the reported moiré superlattices are 2D triangular in shape. Exotic changes in electronic structure due to the superlattice effect have been reported [4,5]. Most recently, the quantum fractal spectrum called Hofstadter's butterfly caused by the interplay between the superlattice potential and magnetic field were observed in graphene on BN substrates [6-8]. Theoretical work has shown that 1D superlattices can generate new zero modes and strong anisotropic effects, which in turn can greatly affect transport and be utilized for lens-less collimation of electrons and other applications [1-3]. To fully explore the tuning capabilities of 1D graphene superlattices, new types of substrates with 1D periodic modulation to graphene are much needed.

High index surfaces can be considered as periodic arrays of atomic steps therefore are natural "superlattice substrates" for quasi-1D periodic modulations. In fact, quasi-1D stripes of monolayer graphite have been successfully synthesized on Ni(771) [24, 25]. The stripes on the terraces have a well-defined width thus exhibiting size limited effects. However for the purpose of the current work on graphene superlattice, large area graphene films continuously extended over the steps are needed.

Because of the high surface energy and confined terrace width, growth of continuous graphene films on high index surfaces has been deemed difficult and is yet to be realized.

As a first demo, we succeeded in preparation of large area monolayer graphene on the high index Cu(410)-O by chemical vapor deposition (CVD) and subsequent annealing. The graphene/Cu(410)-O system was studied by Raman spectroscopy, Auger electron spectroscopy (AES), low energy electron diffraction (LEED), scanning tunneling microscopy (STM) and scanning tunneling spectroscopy (STS). A quasi-1D graphene superlattice that "copies" the form of the Cu(410)-O periodic arrays was observed. Consequently the moiré superlattice is also quasi-1D. New Dirac points generated at the superlattice Brillouin zone (SBZ) boundary were revealed by STS. Similar superlattice-forming phenomena were also observed on Cu(210) and Cu(311) surfaces.

## II. EXPERIMENTS

The graphene samples were grown by CVD on 25 μm thick Cu foils (99.8%, Alfa Aesar) in a quartz tube. The tube was purged with 10 sccm $H_2$ and heated to 1300 K at a pressure of 50 pa. After that 1.1 sccm $CH_4$ was added to the flow and the samples were held at 1300 K and 60 pa for 5~10 minutes. Then the $CH_4$ flow was stopped and the tube was naturally cooled down to below 500 K before the samples were taken out of the tube. The samples were annealed below 650 K in ultrahigh vacuum with base pressure below $3\times10^{-7}$ pa for subsequent AES, LEED and STM measurements.

## III. RESULTS AND DISCUSSION

To check the quality of graphene, Micro-Raman measurements were carried out in ambient conditions (both before and after the ultrahigh vacuum experiments) using a Renishaw RM1000 system with a 514 nm excitation wavelength. Figure 1a is a representative spectrum. The single Lorentz shaped G peak at ~1585 $cm^{-1}$ (with intensity $I_G$) and 2D peak at ~2700 $cm^{-1}$ (with intensity $I_{2D}$), the $I_{2D}/I_G$ ratio of ~4 and the full-width at half-maximum (FWHM) of the 2D peak of ~30 cm$^{-1}$ are good indications of monolayer graphene [26]. Negligibly low D peak indicates low defect density in graphene.

Since exposure to air and annealing in UHV might cause reactions in both graphene and the substrate, AES was performed to check the chemical composition of the samples. A typical result is shown in Figure 1b. The relevant dips correspond to the C KLL line at 271 eV, O KLL line at 503 eV (with intensity $I_O$) and Cu LMM lines around 920 eV (with intensity $I_{Cu}$ at 920 eV), respectively. The oxygen signal remained at the same level after annealing at 600 K in UHV. This amount of oxygen could NOT be due to graphene oxidation giving the negligible D peak in the Raman results. Therefore the oxygen atoms were on the Cu surface below graphene. The AES $I_{Cu}/I_O$ ratio of ~5 is in agreement with the oxygen coverage in the Cu(410)-O phase [27]. The oxygen could come from unavoidable outgas in the CVD furnace and exposure to atmosphere before taking the samples into UHV.

To verify the surface index of the Cu substrate and the relative alignment between graphene and Cu, We carried out LEED measurements. Figure 2a shows a typical result. When increasing the electron energy, the diffraction spots would move towards the (00) beam (specular reflection beam) spot. From this behavior and the symmetry of the diffraction patterns, the (00) beams of graphene and Cu were determined. They coincide in the center of the screen (labeled O), meaning their surfaces parallel to each other (with an error bar of ±2°). The other (00) beam marked by the purple arrow was from another very small grain exhibiting no discernable diffraction patterns, therefore will not be discussed further. Three sets of graphene hexagonal diffraction pattern (corresponding to the A, B and C-labeled spots on the circumference) and a Cu diffraction pattern marked by the yellow arrows were identified. The Cu diffraction pattern persisted even after the graphene layer was removed by annealing at 700K in UHV. A simulation of normal incidence LEED patterns with Cu(410)-O and graphene surfaces parallel to each other is depicted in Figure 2b. The gray spots correspond to the diffraction spots of Cu(410)-O. The spots on the circumference correspond to the three sets of graphene diffraction spots. An excellent match to the experimental result in Figure 2a was identified. Figure 2c-2f are schematic lattice diagrams of the Cu(410) top view, Cu(410)-O top view, Cu(410)-O side view and graphene top view, respectively. The bulk-cut Cu(410) surface may be considered as a periodically stepped vicinal Cu (100) surface with a 7.44 Å step spacing and a 1.75 Å step height from the side view (viewed along the (410) pole). Vectors **a** and **b** are the basic vectors of its primitive cell with an angle of 28° and a = b = 7.66 Å. The ordered Cu(410)-O phase has the same lattice as Cu(410) [27-33]. Vectors **c** and **d** are the basic vectors of the graphene primitive cell. Relevant directions are depicted in Figure 2b-2f. Using the graphene lattice constant of 2.46 Å as a reference, a value of a = 7.5±0.3 Å for the Cu(410)-O was deduced From Figure 2a, close to the pristine 7.66 Å value. The error bar mainly came from the size of the LEED spots. The three sets of the graphene LEED patterns marked A, B, C in Figure 2a and white, green and blue in Figure 2b respectively indicate that there were three predominant azimuthal rotations between graphene [$2\bar{1}\bar{1}$] and Cu(410)-O [001], with "A", "B", "C" sets corresponding to rotational angles of 0°, 21° counter-clockwise and 21° clockwise, respectively (with an error bar of ±2°). This also indicates that there were more than one graphene domains in the area of the incident electron beam (~0.1 mm in diameter).

The formation of large area Cu(410)-O was significantly affected by air exposure after growth and annealing in UHV. This can be understood partly by the fact that many Cu(100) and Cu(111) vicinal facets tend to form {410} facets when exposed to oxygen. Formation of stable Cu(410)-O has been reported on many Cu surfaces upon oxygen dosing, including Cu(511), Cu(610), Cu(711), Cu(11,1,1) and Cu(810) [29-33]. The above Raman, AES and LEED results indicate that the monolayer graphene/Cu(410)-O system can be readily prepared and is reasonably stable. Graphene and Cu(410)-O largely retain their own lattice structures. This point is essential for a "superlattice substrate" to be used as a "mould" for a graphene superlattice.

Having known the geometric corrugation on graphene and the relative alignment between graphene and the Cu substrate, details of the substrate modulation effects on graphene were studied by STM. Representative STM results are shown in Figure 3. Figure 3a and 3b are larger-area scans showing an apparent 1D periodic feature. The bright lines with a spacing of ~7.4 Å (same as the Cu(410)-O step spacing) were observed under a wide range of tunneling conditions (bias −100 mV to −1000 mV, tunneling current 30 pA to 2000 pA). Figure 3c and 3f are atomic resolution zoom-in STM images corresponding to Figure 3a and 3b respectively. While imaging the graphene π state, both the graphene honeycomb lattice (as emphasized by the green hexagon in Figure 3f) and a superlattice consisting of the lines of bright points are clearly seen. Comparing the diamond depicted in Figure 3f with the imaged graphene honeycomb lattice, an angle of 28° and a = b = 7.6±0.2 Å of the superlattice were obtained. The fast Fourier transformation (FFT) pattern corresponding to Figure 3f was presented in Figure 3i. The dots generated by the graphene honeycomb lattice and the superlattice are marked by the green and yellow circles respectively. A similarity between the FFT pattern shown in Figure 3i and the LEED pattern shown in Figure 2a is immediately recognized. As stated above, the LEED pattern was generated by graphene and the underneath Cu(410)-O lattice structures while the STM images reflected the local density of states (LDOS) on graphene surface. Therefore the above similarity demonstrates that the graphene superlattice "copied" the Cu(410)-O super-periodicity.

The rotational angles between the graphene $[2\bar{1}\bar{1}]$ and the Cu [001] were 22º clockwise and counter-clockwise as measured in Figure 3c and 3f respectively, agreeing well with the LEED results. The apparent corrugation in Figure 3c was measured along the dashed line and depicted in the line profile in Figure 3h, given a ~1.5 Å value. The graphene layer smoothly went across not only the periodic [001] steps but also other extra steps of the Cu substrate as shown in Figure 3a and 3b. Sometimes a single graphene domain extended over microns in dimensions. We expect that larger domains can be achieved by optimizing growth conditions.

Besides the above mentioned graphene superlattice, a ribbon-like moiré pattern was also observed as shown in Figure 3c and 3f along the direction of the dotted lines. As a consequence of the 1D modulation by the substrate, this moiré superlattice is also quasi-1D with d = 2.6±0.2 nm, giving additional apparent corrugation of ~0.5 Å to graphene. Geometric simulations of the moiré patterns in Figure 3c and 3f are shown in Figure 3e and 3g respectively. In the FFT pattern in Figure 3i, the moiré feature corresponds to the spot marked by the blue arrow.

To probe the superlattice effect on the electronic structure, STS were taken at 78K. The influence of the 1D superlattice is manifested by the dips at ~±900 meV in Figure 3j. For graphene in a 1D periodic potential, new Dirac points are generated at the SBZ boundary at energies $E = \pm h v_F / 2d$ away from the original Dirac point, where $v_F = 1.1 \times 10^6$ m/s is the Fermi velocity and $d$ is the period of the 1D potential [5, 34]. We note that $E$ depends on the period $d$, not the amplitude or other details of the potential. If we take the 1D moiré spacing in our samples d = 2.6 nm, it gives new Dirac points at ±880 meV which will result in dips in the density of states at these energies, in good agreement with our STS data in Figure 3j. On the other hand, the 7.4 Å superlattice in our samples would generate new Dirac points at energies higher than 3 eV, beyond the stable bias range in our STS experiments thus not detected. Details of the form and magnitude of the superlattice periodic potentials need further study, which can be relevant to other features in the electronic structure.

The hereby reported mechanism of generating graphene superlattices is not limited to Cu(410). For instance, on high index Cu(210) and Cu(311), the graphene superlattice also follows the super-periodicity of the underneath substrate as shown in Figure 4. Transferring graphene to passivated high index semiconductor substrates such as hydrogen-covered Si and Ge should also be possible [35]. This will add a whole new class of quasi-1D "superlattice substrates" to our tuning reservoir and facilitate study on the interesting transport properties of 1D graphene superlattices..

## IV. CONCLUSION

In summary, large area quasi-1D monolayer graphene superlattices on a prototypical high index surface Cu(410)-O have been prepared and characterized by Raman, AES, LEED, STM and STS measurements. One type of superlattice directly follows the super-periodicity of the underneath high index surface. Additionally, this 1D modulation also results in a 1D moiré superlattice with a period of 2.6 nm. The superlattice effect on the electronic structure of graphene is manifested by dips in STS at energies corresponding to new Dirac points generated at the SBZ boundary. This combination of graphene and high index surfaces opens the way to explore interesting properties and potential applications of quasi-1D graphene superlattices.

## ACKNOWLEDGMENT

Z. H. thanks NBRP of China (Grant 2012CB921300), National Natural Science Foundation of China (Grant 11074005 and 91021007) and the Chinese Ministry of Education for financial supports.

**FIGURE CAPTIONS**

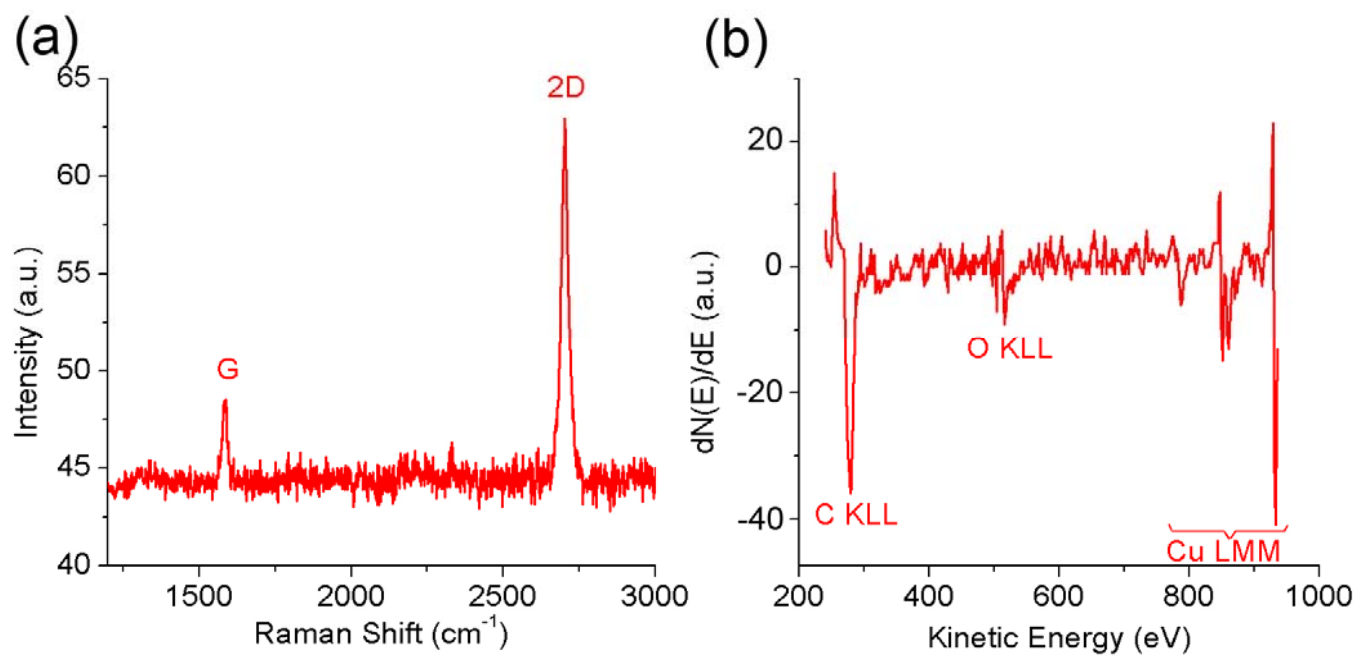

FIG. 1. (a) A Raman spectrum of Graphene on Cu showing the G peak and 2D peak typical of monolayer graphene. The negligibly low D peak at ~1350 cm$^{-1}$ indicates a very low defect density and high quality of the graphene samples. (b) Auger electron spectroscopy data indicating the top layers of the samples mainly composed of C, O and Cu.

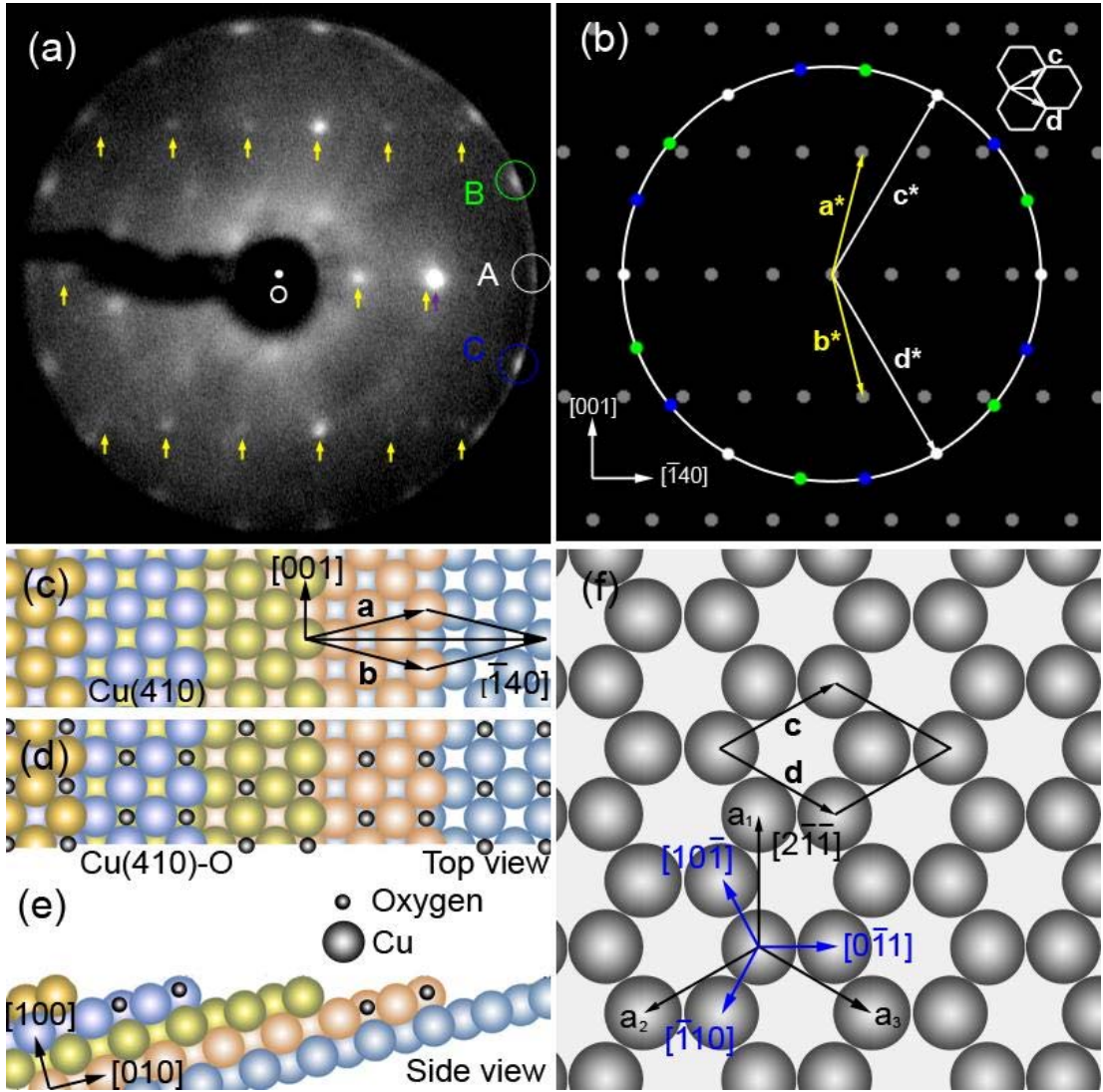

FIG. 2. LEED patterns and schematic diagrams of Cu(410)-O and graphene. (a) A LEED result verifying the relative alignment between graphene and Cu(410)-O. The coincidence of their (00) beams (labled O) indicates their surfaces parallel to each other (with an error bar of ±2°). A, B and C mark three sets of diffraction patterns corresponding to different graphene domains. The yellow arrows mark the Cu(410)-O diffraction spots. The angles of COA and AOB are both 21±2°. (b) A simulation of LEED in (a). The gray dots represent the Cu(410)-O LEED spots. (c) A diagram of Cu(410) top view showing the primitive cell with *a* and *b* as its basic vectors. (d) A diagram of Cu(410)-O top view. (e) A diagram of Cu(410)-O side view. (f) A diagram of the graphene lattice and relevant directions in the two dimensional Miller–Bravais notation ([hki] with $a_1$, $a_2$ and $a_3$ axes for the honeycomb lattice).

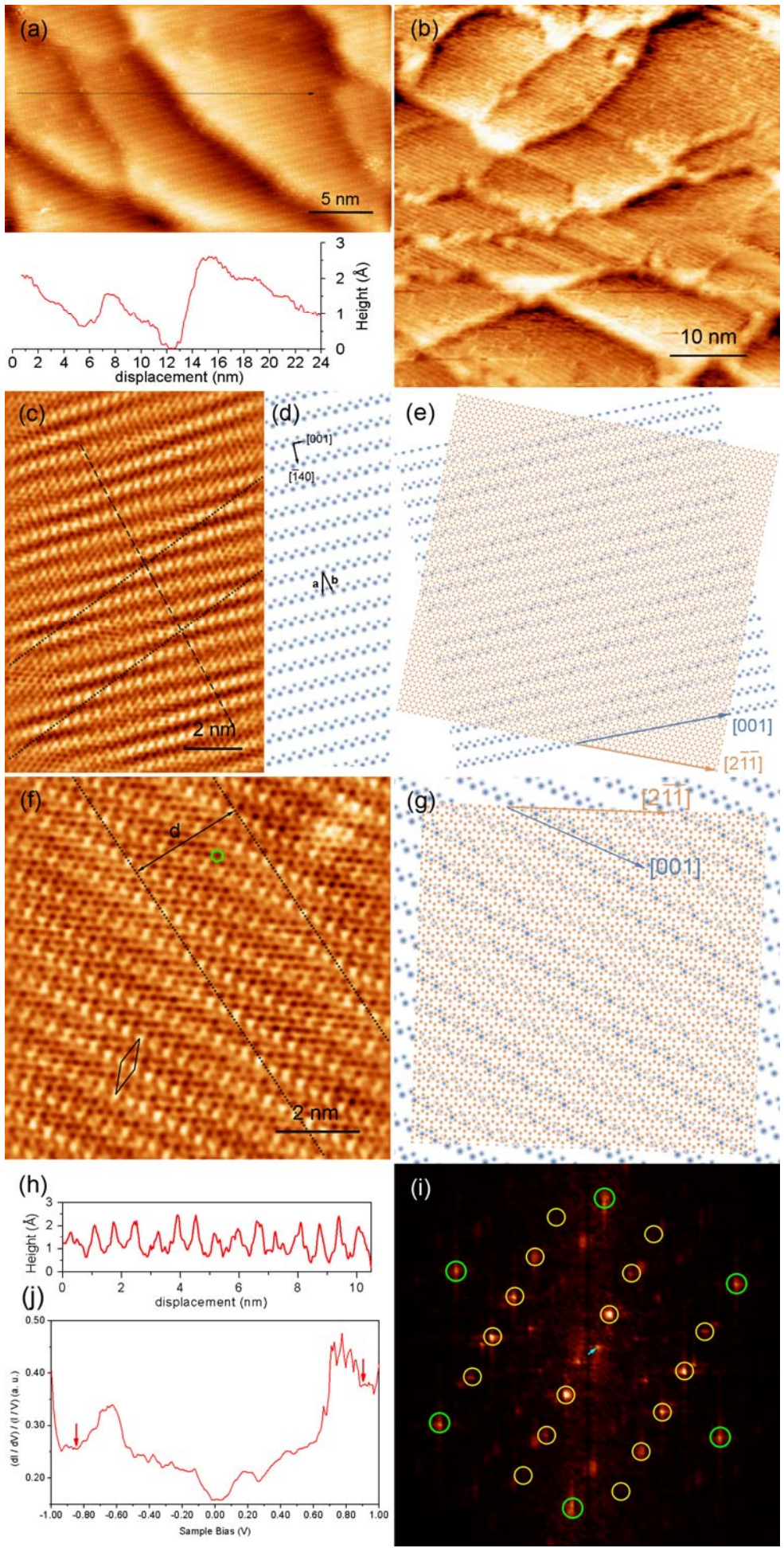

FIG. 3. STM and STS results showing the graphene superlattice directly following the periodicity of the underneath Cu(410)-O super-periodicity and the superlattice potential strongly influencing the electronic structure of graphene. (a) Up: An STM image of graphene on Cu(410)-O showing the large area line-shaped graphene superlattice (scanning conditions: V = -1.0 V, I = 200 pA). Down: Line profile along the arrowed line. (b) An STM image (V = -1.2 V, I = 10 pA) showing another area. (c) An atomic resolution STM image (V = -560 mV, I = 480 pA) corresponding to (a), showing the Cu(410)-O super-periodicity "transferred" to the graphene superlattice and an additional ribbon-like moiré pattern along the direction of the dotted lines. (d) A top view schematic diagram of the top two Cu(410) layers beneath the graphene layer in (c). The large (small) dots represent atoms of the first (second) Cu(410) layer. *a* and *b* are the basic vectors of the Cu(410) lattice. (e) Geometric simulation of (c) using monolayer graphene and top two Cu(410) (which shares the same Bravais lattice with Cu(410)-O) layers. The orange and blue dots represent the C and Cu atoms respectively. Graphene $[2\bar{1}\bar{1}]$ rotates 22° clockwise against Cu [001]. (f) An atomic resolution STM image (V =-1.0 V, I = 40 pA) corresponding to (b). The diamond marks the superlattice cell. The green hexagon marks the graphene honeycomb cell. (g) Geometric simulation of (f). Graphene $[2\bar{1}\bar{1}]$ rotates 22° counter-clockwise against Cu [001]. (h) The line profile along the dashed line in (c) showing the apparent corrugation of the superlattice. (i) The FFT corresponding to (f). The dots generated by the graphene lattice and the superlattice are marked by the green and yellow circles respectively. The other dots are generated by the difference between the vectors corresponding to the green and yellow circles, inducing the moiré pattern shown in (f). (j) STS data taken at 78K showing new dips at about ±900 meV marked by the arrows, corresponding to newly generated Dirac points by the quasi-1D 2.6 nm moiré superlattice.

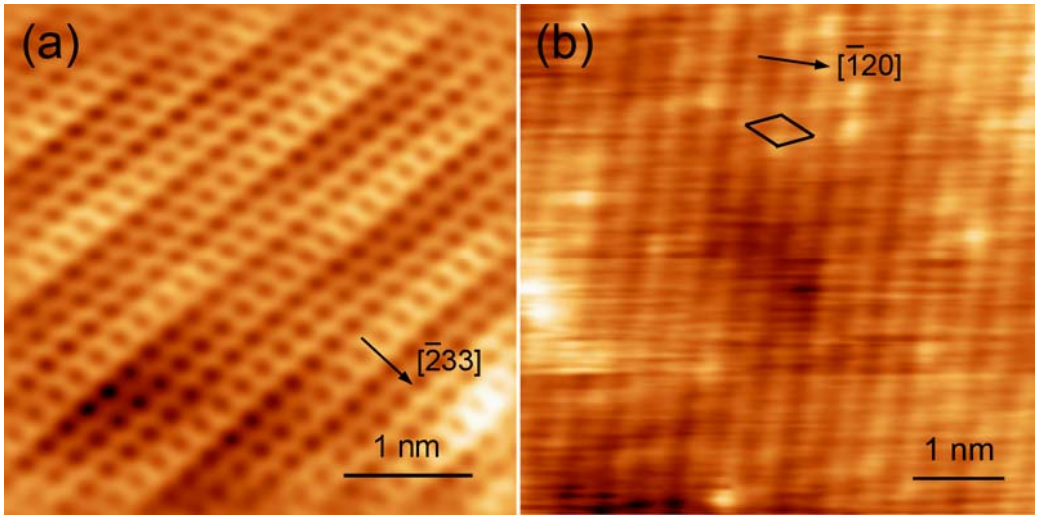

FIG. 4. More examples of "pattern transfer" from high index surfaces (superlattice substrates) to graphene superlattices. (a) An STM image of quasi-1D graphene superlattice on Cu(311) (V = -0.55 V, I = 0.80 nA). The spacing (marked by the arrow) between the 1D lines is 0.42 nm, same as the Cu(311) step width along [$\bar{2}$33]. (b) An STM image of quasi-1D graphene superlattice on Cu(210) (V = 0.50 V, I = 1.15 nA). The spacing (marked by the arrow) between the 1D lines is 0.40 nm, same as the Cu(210) step width along [$\bar{1}$20]. The diamond marks the Cu(210) primitive cell.